\documentclass[article]{aa}  

\usepackage{graphicx}
\usepackage{txfonts}
\usepackage{hyperref}
\hypersetup{colorlinks=true, allcolors=blue}
\begin{document}

   \title{Requiem for a belt}

   \subtitle{A spatial and kinematical reinterpretation of Gould's Belt in light of \textit{Gaia}}

   \author{
   M. Pantaleoni Gonz\'alez,\inst{1}
   J. Alves,\inst{1}
   C. Swiggum,\inst{1,2}
   \and
   I. Niederbrunner \inst{1,3,4}
   }

   \institute{
   University of Vienna, Department of Astrophysics, T\"urkenschanzstrasse 17, 1180 Wien, Austria\\
   \email{michelangelo.pantaleoni@univie.ac.at}
   \and
   Center for Astrophysics, Harvard \& Smithsonian, 60 Garden St., Cambridge, MA 02138, United States of America
   \and
   Departament de Física Quàntica i Astrofísica (FQA), Universitat de Barcelona (UB), Martí i Franquès, 1, 08028 Barcelona, Spain
   \and
   Institut de Ciències del Cosmos (ICCUB), Universitat de Barcelona (UB), Martí i Franquès, 1, 08028 Barcelona, Spain}

   \date{Received XXXX XX, 2026; accepted XXXX XX, 2026}

 
  \abstract
   {We reassess the long-standing idea of Gould's Belt using the third \textit{Gaia} data release (DR3) for a sample of young massive stars and nearby young clusters. The structure surrounding the Sun, often interpreted as an inclined, expanding, and rotating ring, emerges in our analysis as a transient alignment of a few cluster families rather than an individual, coherent dynamical feature. By combining the ALS III catalog of OB stars with a homogeneous sample of clusters younger than $70$ Myr, and by tracing their motions in a realistic Galactic potential, we show that neither the spatial distribution nor the kinematics form a unified system. The inferred expansion, rotation, and bulk motion of the Belt can be reproduced by the superposition of the $\alpha$Per, Cr135, M6, and $\gamma$Vel cluster families and are further amplified by solar reflex motion and historical assumptions about the local standard of rest (LSR). The classic inclined geometry is largely explained by the oscillatory pattern of the Radcliffe Wave, which contributes a major arc of the supposed ring. Taken together, these results indicate that Gould's Belt is not a physical structure but a 3D asterism shaped by a complex local star formation history, observational biases, and projection effects.}
   



   \maketitle
%

\section{Introduction}

During the last 170 years, astronomers have debated about the existence, origin and properties of an evasive structure known as ``Gould's Belt''. It was first noted by \citet{Hers47} as a band of bright stars visible in the southern skies (possibly inspired by his father's concept of a ``\textit{secondary branch}'' of the ``\textit{main sidereal stratum}'' seen in the northern hemisphere; \citealt{Hers84}). It was soon realized that this feature followed a great circle around the entire celestial sphere, inclined by some $\sim20^{\circ}$ with respect to the Galactic plane \citep{Stru47, Goul74, Goul79} and that it was probably constituted by the nearest bright stars \citep{Celo77}. Despite this, already in \citet{Newc04} suspicions were raised in regards to the true cosmographical significance of the Belt, with the suggestion that it might be an accidental disposition of nearby stellar aggregates, particularly highlighted by those in Orion (a view later supported by the analysis in \citealt{Pann29}). Nonetheless, other studies made clear that the Belt was traced by dark clouds and reflection nebulae \citep{Hubb22, van66},  A-type \citep{ShapCann22a} and B-type stars \citep{ShapCann22, ShapCann24}. For over a century, various models, such as flat disks \citep{Shap18, Guiletal98}, rings \citep{StotFrog74, Olan82}, and even rectangular cuboids \citep{Club67} were proposed with the hope of characterizing the tridimensional shape of this feature. By the second-half of the 20th century an associated ring of HI showing signs of expansion had been detected \citep{Lind67}, and molecular clouds were been used as tracers of the Belt \citep{Tayletal87}, aswell as late-type stars \citep{Guiletal98}. Gould's Belt was subsequently said to be expanding \citep{Club67, Lesh68}, rotating \citep{Olan82} and translating as a whole \citep{Boby04}. With the advent of space-based astrometry and the completion of the \textit{Hipparcos} mission, it became possible to use 3D positions and kinematics to investigate the nature of this structure. \citet{Lindetal97} analysed what was then the most accurate 3D map of OB stars in the solar vicinity. The characteristic inclined plane of Gould's Belt was clearly visible, but no ring-like structure appeared in the top-down view of the Galactic plane. Further analysis strengthened the view that the Gould's Belt was a genuine physical structure that could be characterized with a simple elliptical ring model, with a specific size, inclination, and orientation within the plane of the ellipse \citep{PerrGren03}. The possible formation scenarios explored included violent gas expansion due to a series of supernovae \citep{Olan82, Popp97}, the gradual acceleration due to stellar winds or ionization-front driven pressure waves \citep{Tayletal87}, the natural evolution of a supercloud that encountered a density wave \citep{Olan01}, the passage of a globular cluster in this region of the disk \citep{BobyBajk18a}, the high velocity collision with an intergalactic cloud \citep{Comeetal92} and even a speculative encounter with a dark matter clump \citep{Bekk09}. The expansion rates obtained using different tracers allowed for age estimates ranging from $\sim 30$ Myr to $\sim 90$ Myr \citep{Comeetal92, Lindetal97}. Hopes of reaching a final consensus on the matter were renewed with the announcement of ESA's \textit{Gaia} mission \citep{Prusetal16, Valletal23}, which was expected to deliver micro-arcsecond astrometry and an unprecedentedly clear view of the first kiloparsec around the Sun. For a detailed account of the historical debate surrounding Gould's Belt prior to the advent of \textit{Gaia}, see the reviews by \citet{Popp97} and \citet{PaloEhle17}.

Despite accumulating evidence, skepticism about the existence of the Belt persisted. The map in \citet{Lindetal97} had failed to show a ring-like structure around the Sun, and the observed kinematics were incoherent enough to be attributed to unrelated star formation episodes \citep{Eliaetal09}. When \citet{Moreetal99} calculated the deviation of the vertex, $l_v$, for a sample of stars in Gould's Belt, they found a significant negative value, against the predictions of basic Galactic dynamics. But when the stars in the Pleiades moving group (which can be said to overlap with the $\alpha$Per cluster family; \citealt{Swigetal25}) were removed from their sample the deviation of the vertex returned to a positive value, strongly suggesting that the full sample was not representative of a single unified structure but that it was in fact a patchwork of different stellar populations. An alternative view suggested that the Belt could simply be a projection effect created by the distribution of nearby OB associations shown by Hipparcos data \citep{BouyAlve15}. Recently, in the post-Gaia era, the discovery of the Split \citep{Lalletal19} and the Radcliffe Wave, two parallel linear structures each containing the main stellar populations of the Belt (Sco-Cen and Orion), gave further support to this interpretation \citep{Alveetal20}. Hopes that the \textit{Gaia} mission's unparalleled astrometric precision would clarify the structure were dashed, as recent 3D dust maps (\citealt{Greeetal19, Lalletal19, Leiketal20, Vergetal22, Edenetal24}) and the distribution of OBA stars \citep{Zarietal18, Poggetal21, Zarietal21} showed no evidence of the purported Belt. On top of that, all the properties and structural parameters previously attributed to this structure seem to be largely conditioned by just two prominent associations in the Belt: Orion OB1 and Sco–Cen OB2 (\citealt{Alfaetal22}). But even if the current scientific consensus leans toward interpreting Gould's Belt as a mere pareidolia in six-dimensional phase space, the topic continues to generate substantial debate, and the hypothesis of its existence continues to be defended by some authors \citep{Dzibetal18, McBretal21, Gontetal22, Bobyetal25}.

\begin{figure*}
    \centering
    \includegraphics[width=1.0\textwidth, trim=50 25 50 25, clip]{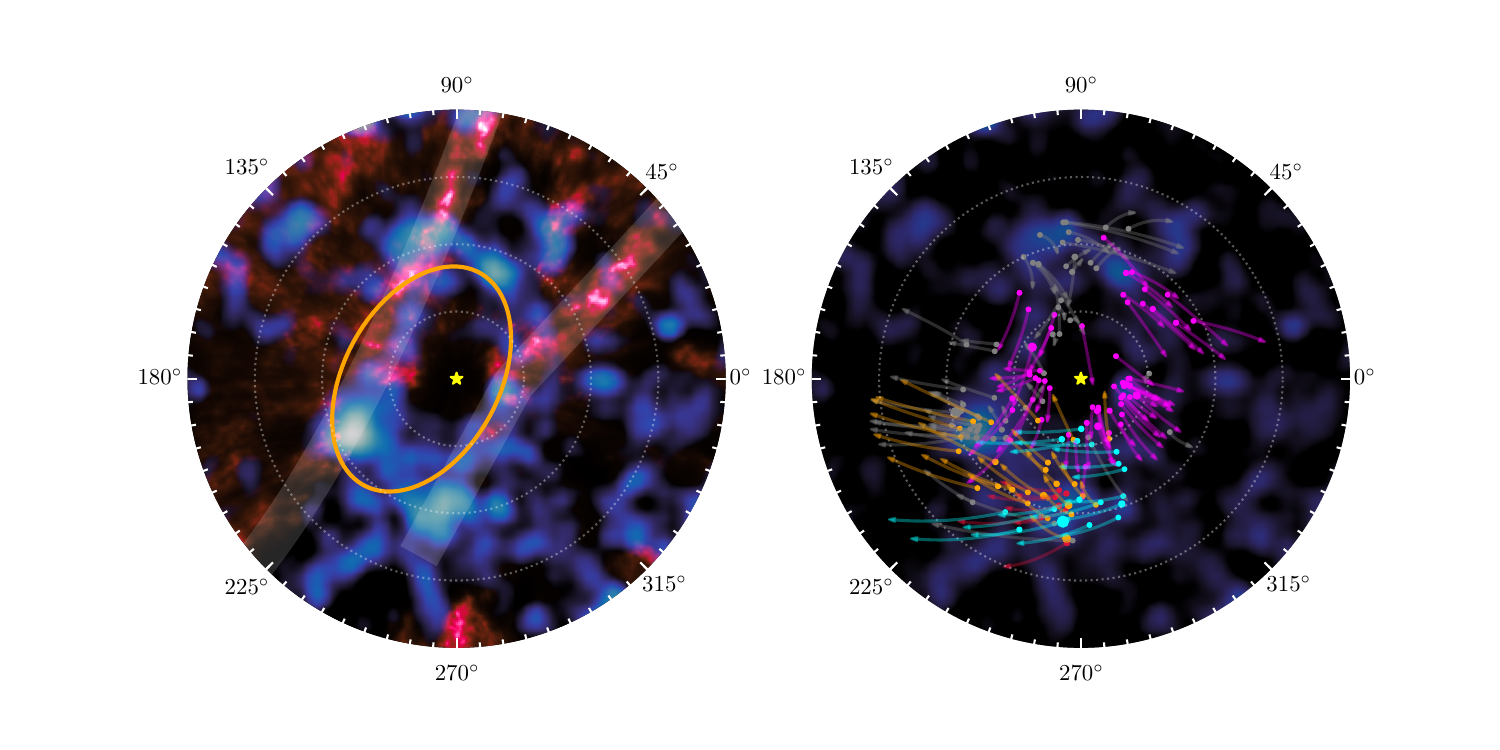}
    \caption{View from the north Galactic pole of the $800$ pc solar neighbourhood with the Galactic Centre towards the right. A set of dashed concentric circles are placed each $200$ pc from the Sun, which is itself represented by a yellow star. (\textit{Left panel}) An overlay of the OB star density field from \cite{Pantetal25} (blue) and the dust distribution from the extinction map in \citet{Vergetal22} (red). The Radcliffe Wave (left; using the best-fit model in \citealt{Konietal24}) and the Split (right) are shown as lightly shaded regions, while the Gould's Belt reference model from \citet{PerrGren03} is shown as an orange ellipse. (\textit{Right panel}) Star clusters from \citet{HuntReff23}, associated with Gould's Belt, together with their expected trajectories for the next $15$ Myr in the frame of the LSR. Colours indicate the different cluster families in \cite{Swigetal24}; $\alpha$Per (magenta), M6 (cyan), Cr135 (orange), in addition to the small $\gamma$Vel family (crimson) and non-grouped young clusters (grey), many of which are members of the Radcliffe Wave. Marker sizes are proportional to the masses of the clusters.}
    \label{fig:fig1}
\end{figure*}

\section{Data \& Methods}

To characterize the putative belt, we use samples of OB stars and young clusters. For the stellar component, we adopt the young massive star catalogue from \cite{Pantetal25} (hereafter ALS III), selecting sources assigned to the "M" category. These are stars confirmed as massive OB through a combination of high-quality astrophotometry and literature classifications. We further restrict the sample to stars located within $0.75$ kpc of the Galactic mid-plane, in order to minimize contamination from the thick disk and halo populations (typically composed of young massive runaways). In the ALS III catalogue, OB stars are defined strictly as those with masses exceeding $8M_{\odot}$, typically corresponding to spectral types of B2 or earlier for the main sequence, B5 or earlier for the giants, and extending to spectral types as late as A-type stars for the supergiants (see \citealt{Maizetal26}).

To investigate the properties of Gould's Belt, we select all stars and clusters located within $150$ pc of its outline, which we defined with an ellipse approximating the distribution shown by the OB stars. This yields a sample of 338 young massive stars. Of these OB stars, $57\%$ have been confirmed as such by ground-based multi-epoch spectroscopy, while the remaining stars are classified as massive in accordance to previous literature and satisfying the ALS III criteria based on \textit{Gaia} DR3 photometry and distances (placing them consistently above the $T_{\text{eff}} = 20$ kK extinction track in the Hertzsprung–Russell diagram).

The cluster sample is taken from the catalogue in \cite{HuntReff23}, which used HDBSCAN to accurately identify nearby clusters in \textit{Gaia} DR3. We then select those clusters younger than $70$ Myr, leaving $160$ of them within the confines of the Belt. Following the grouping scheme proposed by \citet{Swigetal24}, these clusters are distributed among three major cluster families: $\alpha$Per (57 members within our Belt selection), Cr135 (25 clusters), and M6 (17 clusters). In addition, we identify a smaller but coherent grouping, the $\gamma$Vel family, comprising seven additional clusters. The remaining 54 clusters (comprising a third of the cluster sample) do not belong to any of these families, albeit most of them are associated to the Radcliffe Wave.

Both the young stellar and young cluster Gould's Belt samples are based on high-quality \textit{Gaia} DR3 astrometric and photometric data. For the stars, proper motions have been corrected following \citet{CantBran21}. Also, parallax zero-point adjustments from \citet{Maiz22} were applied before estimating their distances with a bayesian approach. A detailed discussion of these corrections, as well as of the treatment of associated uncertainties (which are underestimated in \textit{Gaia} DR3; \citealt{Maizetal21a, Maiz22, El-B25}), is provided in the ALS III paper. Distances to individual stars are taken as the median of the posterior distributions derived in the ALS III. For clusters, these distances are computed from the weighted averages of the individual stellar parallaxes, which allows higher levels of accuracy (consequently also in tangential speeds) without the need of additional refinements.

For the kinematical analysis of Gould's Belt clusters (Section \ref{sec:results}), we perform back-tracing calculations through the galpy \texttt{MWPotential2014} model of the Galactic potential (\citealt{Bovy15}), which includes a Galactic halo, bulge, and disk. Although the model neglects non-axisymmetric structures like spiral arms, its accuracy is sufficient for our purposes, as the integrations extend no further than $45$ Myr in time \citep{Arunetal25}. Unless noted otherwise, we define the Local Standard of Rest (LSR) using the Sun's peculiar motion of ($U_{\odot}$, $V_{\odot}$, $W_{\odot}$) $=$ ($11.1$, $12.24$, $7.25$) $\pm$ ($0.75$, $0.47$, $0.37$) km/s, determined by \citet{Schoetal10}. The Galactic potential is scaled in order to match a LSR circular velocity of $V_c = 236 \pm 7$ km/s \citep{Reidetal19} for a Galactocentric distance of $R_0 = 8.178\pm0.026$ kpc \citep{Abutetal19}.

For Section \ref{sec:copernicus}, we also compute the peculiar velocities of our stellar and cluster samples using the rotation curve from \citet{Sofu20} (our results are not particularly sensitive to the choice of rotation curve since all tend to agree on the fact that it is flat in the region considered in this paper). For massive stars, we limit the calculation to tangential peculiar velocities (in contrast to the total velocities computed for the young clusters), since radial velocities are generally unavailable for this population, and when they are these measurements can be highly unreliable due to biases related to high multiplicity and the strong stellar winds, ubiquitous in OB stars. We also compute the peculiar velocities for an ``outside-the-Belt'' sample of OB stars and clusters, defined as those with distances between $600$ pc and $1200$ pc. The massive star sample and cluster samples outside the Belt contain $1403$ stars and $45$ clusters, respectively.

\section{Results: A Revised View of Gould's Belt}
\label{sec:results}

Gould's Belt is currently said to be characterized by \textbf{1)} a $\sim0.7 \times0.5$ kpc elliptical ring of young stars, dust and gas \citep{Olan82, PerrGren03}, centered at just $\sim 80-150$ pc from the current position of the Sun towards the $l\sim 150^\circ$ direction \citep{Olan82, ComeTorr91, Boby04}, inclined by some $i \sim 15^\circ-20^\circ$ with respect to the Galactic plane \citep{Torretal00, Boby04, Eliaetal06}, that is, \textbf{2)} expanding \citep{Lindetal73, Olan82} at a rate of $2.5\pm0.1$ km s$^{-1}$ \citep{Dzibetal18}, and \textbf{3)} rotating with an angular velocity of $\omega \simeq 24$ km s$^{-1}$ kpc$^{-1}$ \citep{Lind00}, while having a \textbf{4)} bulk translational motion relative to the LSR of $\sim 10$ km/s towards $l \sim 270^\circ$ \citep{Boby04, Boby06}. In the following, we critically revisit these properties and offer a reinterpretation in light of the new cosmographic developments brought by the \textit{Gaia} era.

\begin{figure*}
    \centering
    \includegraphics[width=1.0\textwidth, trim=60 5 60 25, clip]{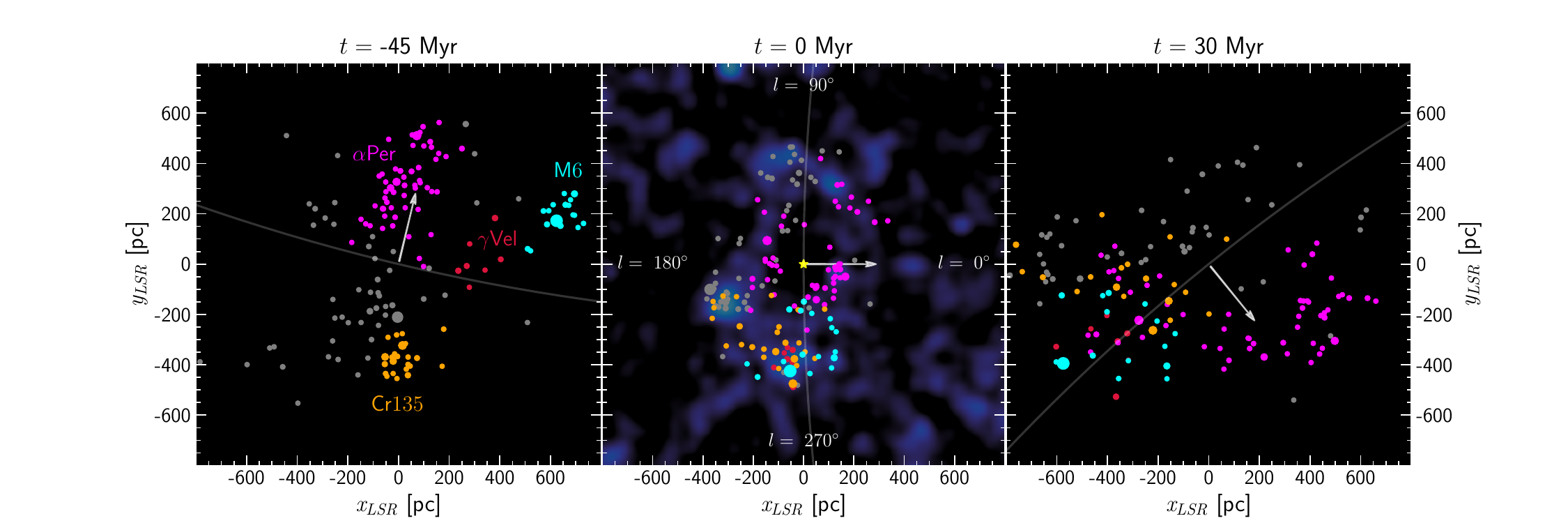}
    \caption{Top-down LSR-centered views showing the evolution in the distribution of the young cluster sample associated with Gould's Belt: (\textit{Left panel}) $45$ Myr ago, (\textit{central panel}) present day, and (\textit{right panel}) $30$ Myr into the future. The central panel also shows the OB stars depicted in Fig. \ref{fig:fig1}. Clusters are coloured according to the families. The orientation of the frame of the LSR remains fixed across panels. The grey curve is an arc of the solar circle and the grey arrow in each panel guides the viewer toward the Galactic Centre. The yellow star marking the Sun falls outside the plotted area in both the left and right panels, as its peculiar velocity far exceeds the modest velocities of the young clusters. Note that the cluster families were more compact at the time of their individual formation than is suggested by the left panel. However, due to the slight differences in their formation times, we adopt $45$ Myr in the past as a common reference epoch to illustrate their overall compactness.}
    \label{fig:fig2}
\end{figure*}

\subsection{Spatial distribution: Ring-like structure and inclination}

Starting from the direction of the Galactic centre and following decreasing longitude (clockwise in Fig. \ref{fig:fig1}), the ring associated to the Belt is delineated by the young stellar associations Sco–Cen OB2, Vel OB2, Ori OB1, Per OB2, Cep OB6 (Cepheus–Near), Lac OB1, and finally Cep–Her (see \citealt{Wrig20, Kerretal24, Quinetal25}). Among these, Ori OB1 (notably located below the mid-Galactic plane) and Sco–Cen OB2 (above the plane) have long been regarded as the most prominent members of the Belt \citep{deZetal99}. Their contrasting positions were historically taken as key evidence that the structure lies on an inclined plane. But, with the identification of the Radcliffe Wave as a single coherent structure, this interpretation has shifted, as a segment of the Radcliffe Wave is now considered to be the anti-central arc of the Belt (from Ori OB1 to Lac OB1). In this view, the anomalous height of Ori OB1 is naturally accounted for by the nearest most visible portion of the oscillatory pattern attributed to the Radcliffe Wave, rather than by any connection to Sco–Cen. On the other hand, Sco-Cen follows a separate history as a member of the $\alpha$Per family. This interpretation is also supported by the contrast in age of the different arcs of the Belt (see Fig. 12 in \citealt{Kerretal23}).

Recent studies reveal that the dust does not trace a continuous ring in the same way the young stars in the Belt do but are instead arranged in parallel structures \citep{Edenetal24, Kormetal25}. The Radcliffe Wave and the Split (which coincides with the Sco-Cen - Vel OB2 arc of the Belt) form the dust lanes comprising the Orion, Taurus and Perseus clouds on one hand and the Aquila Rift on the other. Between these two lanes there are two bridges of OB stars (including Cep-Her; see \citealt{Kerretal24}) closing the circuit. But these connecting arcs appear only faintly (or not at all) in the dust. This explains why, between experts on the local ISM, the Gould's Belt hypothesis has slowly been replaced by the idea that the Belt is a mere projection effect on the celestial sphere, created by the Radcliffe Wave and the Split.

Besides a small shift of a few tens of parsecs towards the Galactic center, the most reliable model for the Gould's Belt geometry \citep{PerrGren03} matches the properties of the spatial distribution of nearby OB stars (left panel in Fig. \ref{fig:fig1}). We argue that the ``projection hypothesis'' would therefore work only for the dust component, and claim that there is indeed a tridimensional ring of young massive stars surrounding the Sun. While future observations will refine individual distances, we do not expect the overall spatial pattern to change appreciably, as the large improvement in accuracy from \textit{Gaia} DR2 \citep{Pantetal21} to \textit{Gaia} DR3 (ALS III) already brought only minor adjustments to this area of the Galactic map (having average distance uncertainties of less than $40$ pc). Since Gould's Belt is traced mainly by stars earlier than B2.5 \citep{Popp01}, its non-detection in previous \textit{Gaia}-based studies of Galactic ``OB'' stars \citep{Poggetal21, Zarietal21} likely reflects a combination of inaccurate spectral classifications and broader sample definitions that include non-massive hot stars. These kinematically-hot populations dominate the samples and wash out the contrast of genuine sub-kpc overdensities in the young local disk (for additional details, see Section 2.1 of the ALS III paper). In contrast, studies adopting stricter definitions of what constitutes a young massive star do reveal a ring-like structure in the solar vicinity (see Fig. 1a in \citealt{Moreetal99} and Fig. 12 in \citealt{Kerretal23}). In section \ref{sec:copernicus}, we will also discuss common biases that could overemphasize the apparent ring-like distribution of nearby stars and may account for the persistence of this structure in the literature prior to the advent of micro-arcsecond astrometry.

\subsection{Kinematics: rotation, expansion and translation}

While the spatial arrangement of young stars satisfies the classical expectations for a Gould's Belt, the kinematics do not. The motions of these stars reveal that the apparent ring is the product of a fortuitous alignment rather than a physically meaningful, dynamically coherent structure. Essential to this picture is the recent discovery that many of the nearby young clusters are grouped into three massive and one smaller cluster families \citep{Swigetal24}. Each family being a grouping of clusters that formed together in the same complex. When Gould's Belt was first traced using nearby young clusters \citep{Pisketal06}, the idea of dozens of independent clusters forming a seemingly ordered arrangement seemed improbable. However, if we view the ring as composed of four cluster families (largely independent form each other), the likelihood that its ring-like appearance arises from a chance disposition of these parts becomes more acceptable. All previous models for the formation of the expanding Gould's Belt posit a ring of smaller size in the past. In contrast, as can be seen in Fig. \ref{fig:fig2}, these families formed more compact groupings around $45$ Myr ago (close to the times of their respective formation). More importantly, the member clusters didn't trace a ring, and were distributed in a larger volume (albeit in a more heterogeneous way), which contradicts the notion of a radially expanding structure. Notably the $\alpha$Per family is nowadays distributed bimodaly between the Radcliffe Wave and the Split, as the Sco-Cen clusters detached from the rest of the family and moved across the region now occupied by the Sun. During this time the Cr135 family expanded into the bridge connecting Ori OB1 and Vela OB2. Vela OB2 itself is in fact a combination of the clusters of the M6 and $\gamma$Vel families, which moved quickly from the inside of the Galaxy. Finally, we can see that in less than $30$ Myr into the future, all these clusters will scatter leaving no resemblance of a ring, which strongly implies the ephemeral nature of Gould's Belt.

When analysing the residual motions of nearby young stars (ages $\lesssim 30$ Myr) located between $100$ and $600$ pc from the Sun (roughly the region associated with Gould's Belt), \citet{Torretal00} derived an expansion parameter of $K = 7.1 \pm 1.4$ km/s/kpc, in good agreement with the earlier estimate of $K = 7.4 \pm 2.7$ km/s/kpc by \citet{West85}. These values are smaller, but still broadly consistent with the $K = 11.0 \pm 3.5$ km/s/kpc reported in \citet{Lindetal97}. As stated before, the expansion of the Belt prompted decades of speculation about the local star formation history. But serious doubts about a coherent expansion have long persisted. \citet{Lindetal97} demonstrated that the young stars could not have expanded from a small area in the field of Galactic differential rotation because this would have resulted in the nullification of the second Oort constant, which is certainly contradicted by the observations. By seeking to preserve an expansion scenario, they hypothesized that the entire progenitor cloud of Gould's Belt could have been rapidly rotating, beyond the limits of its self-gravitational binding, and thereby was already expanding before the stars were born. Then in \citep{Boby04} it was shown that for the overwhelming majority of nearby OB stars the $K$ parameter was actually negative and thus a contracting component had to be considered in order to filter this population from the young stars in the ``true'' expanding Belt. A careful analysis by \citet{Eliaetal06a} showed a much smaller value for $K$ than previous studies, discouraging further conclusions about the possible expansion movements of the system using this metric. Critically, \citet{Alfaetal09} showed that even if the characterization of Gould's Belt were reduced to just two representative points (namely the centroids of the main nearby associations, Ori OB1 and Sco–Cen OB2, together with their average velocities) we would still recover the values for the inclination, the longitude of the ascending node, the expansion velocity, and even the residual rotation usually attributed to Gould's Belt, strongly implying that the whole idea of its existence was actually resting on the notion that Orion and Sco-Cen had to be somehow related.

We can now follow the full implications of this interpretation in light of our knowledge of the cluster families and their dynamics. The two most active and massive associations in the local $0.5$ kpc have histories that are largely independent of each other; The older ($>10$ Myr) clusters in Ori OB1 are members of the Cr135 family (suggesting a link between the present-day Orion and past star-formation of the Cr135 family). Sco-Cen on the other hand is a portion of the $\alpha$Per family that separated and accelerated towards the inner Galaxy while inducing the formation of the Local Bubble \citep{Benietal02, BergBrei02, Zucketal22, Ratzetal23, ONeetal24}. In the right panel of Fig. \ref{fig:fig1} we can see how the motion of Ori OB1 towards the anticentre and the motion of Sco-Cen towards the the third Galactic quadrant could convey the idea of an expansion and a torque just by themselves. This illusion is strengthen by the motion of the M6 and $\gamma$Vel families that in this moment in time are contributing to this non-orchestrated choreography in the region occupied by Vel OB2. We can also see that the clusters in Lac OB1, stand in defiance of this picture of an expanding and rotating ring, as they seem to move towards us and would be following counter-rotating paths (together with the clusters in Per OB2 and Per OB3). Lac OB1, however, poses little challenge to the supporters of the Gould's Belt hypothesis, largely because it had already been set aside as an inconvenient case \citep{PerrGren03}. Although traditionally included in the Belt \citep{Olan82}, subsequent analyses showed that its location would require the introduction of an artificial ``corner'' into the otherwise simple elliptical model, and that its dynamical age would be incompatible with the Belt's proposed expansion \citep{deZetal99}. The boundary between rejecting data for not contributing to a specific pattern, on the basis that it represents diverging behavior, and plain cherry-picking is an epistemological issue of relevance when dealing with complex systems like the local Milky Way. If strengthening the signal of Gould's Belt requires the continued removal of data points without alternative justification, the core hypothesis may no longer be tenable.

Finally, a bulk translational motion of $10.7 \pm 0.7$ km/s towards the $(l,b) = (274^\circ, -1^\circ) \pm (4^\circ, 3^\circ)$ direction was also proposed for Gould's Belt in \citet{Boby06}. At the time, this offered further support for the interpretation of the Belt as a genuinely cohesive system that can be discriminated from the overall local Galactic population. However, interpreting the Belt's kinematics in this aspect is challenging, as the solar reflex motion induces a signal in nearby stars that can be easily mistaken for a coherent kinematic pattern. We will provide further details in regards to these biases in section \ref{sec:copernicus}. We don't see any significant residual average global motion of the clusters in the Belt when using the LSR defined by \citet{Schoetal10} (something that can be quickly deduced by the lack of any significant drift as we follow the LSR across the panels in Fig. \ref{fig:fig2}).

\section{Why the Gould's Belt hypothesis persisted}
\label{sec:copernicus}

Challenging a long-standing hypothesis requires not only the proposal of an adequate alternative (Section~\ref{sec:results}), but also an explanation of how previous analyses may have contributed to its apparent validity. In the case of Gould's Belt, several of its defining features have been already shown to be prone to misinterpretation. Rather than attempting an exhaustive review of the literature, we highlight here a set of recurring biases that can arise when interpreting the nature of structures in the solar neighbourhood. While the issues presented in this section are not ubiquitous in the literature regarding Gould's Belt, they appear with sufficient frequency to warrant careful consideration, and must be taken into account to ensure that future discussions do not repeat superseded arguments. In the case of Gould's Belt we are faced with two Copernican cautionary tales: a reminder to treat with skepticism any structure that appears to single out the observer's location, and whose kinematics seem to be suspiciously correlated with those of the observer.

\subsection{Biases leading to apparent ring-like structures}

\begin{figure*}
    \centering
    \includegraphics[width=0.9\textwidth, trim=50 10 30 25, clip]{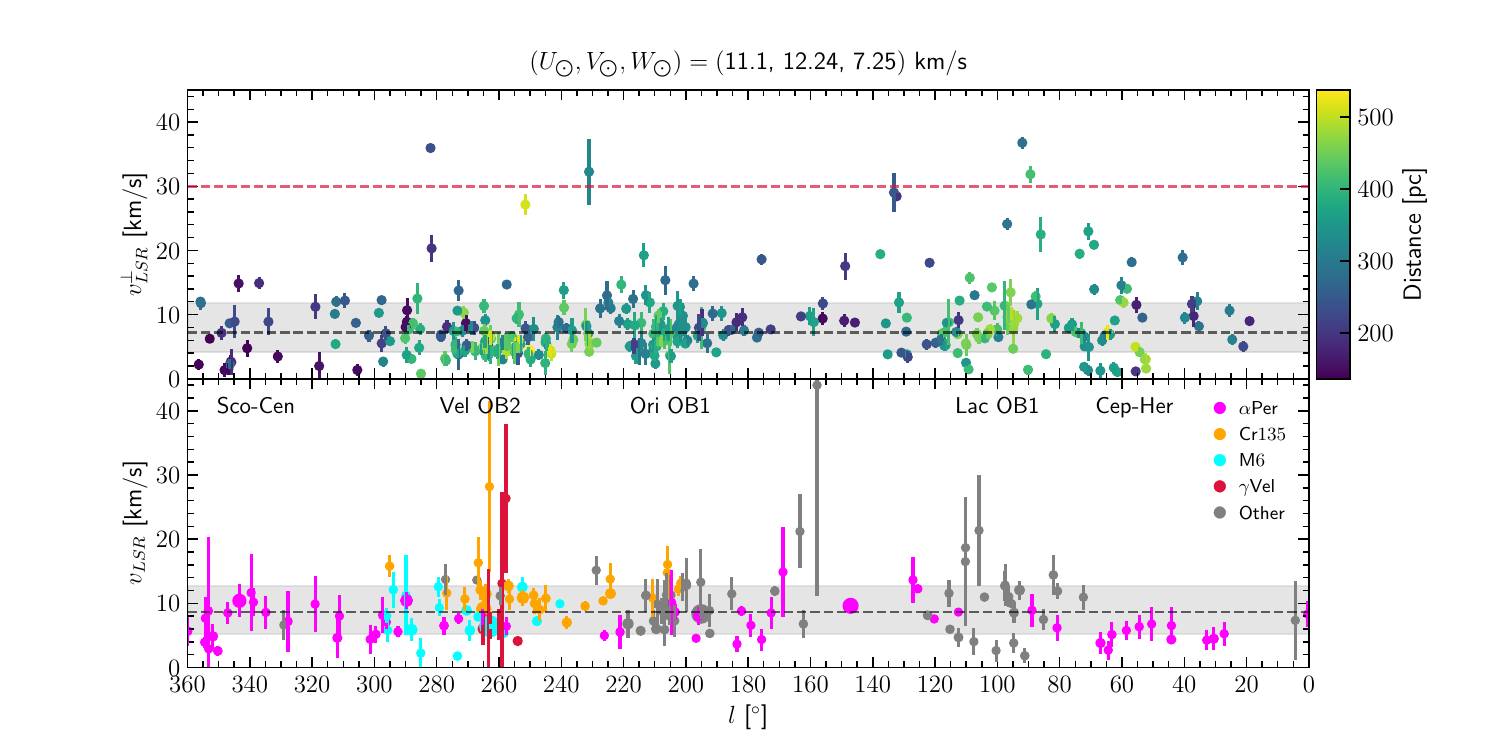}
    \caption{(\textit{Top panel}) Peculiar tangential velocities for the OB stars in Gould's Belt across Galactic longitude, with individual distances shown as a colour gradient. We adopt the peculiar tangential velocity rather than the total peculiar velocity because the latter requires radial velocities, which are largely unavailable for OB stars in \textit{Gaia} DR3 and, when present, are often unreliable due to the high multiplicity and strong stellar winds associated with this population. The red dashed line denotes the $30$ km/s threshold typically used to classify stars as runaways \citep{Blaa61}. (\textit{Bottom panel}) Total peculiar velocities for the clusters (color-coded according to the cluster families and shown with marker sizes proportional to their mass). In both panels, black dashed lines mark the sample medians, and the grey shaded regions trace the range enclosed by the 16th and 84th percentiles. All peculiar velocities are computed with respect to the LSR defined by \citet{Schoetal10}. An \href{https://youtu.be/LSYs4zPPMDQ}{online animated version} of this figure illustrates the effect of varying $V_{\odot}$, showing that the peculiar velocities are minimized near the adopted values, both for the OB stars and the young clusters, leaving no clear signature of a peculiar Gould’s Belt motion.}
    \label{fig:fig3}
\end{figure*}

In this paper, we argue that the number density of young stars increases rapidly with distance beyond the boundary of the Local Bubble, and subsequently declines until the Cepheus and Sagittarius spurs \citep{Kuhnetal21} are reached at distances of $\sim 1$ kpc. However, this ring-like distribution was not clearly discernible prior to \textit{Gaia} and the advent of modern OB star catalogues. Earlier studies nevertheless reported a ring centered on, or near, the Sun. While this interpretation is partly justified by the underlying structure itself, its apparent prominence was likely amplified by a combination of well-known observational biases that merit our attention.

The Malmquist bias \citep{Malm22, Malm25} arises from the use of a limiting magnitude in astronomical surveys: at larger distances, only intrinsically brighter objects remain detectable, leading to a distance-dependent over-representation of luminous stars. Even when restricting the sample to a given luminosity class through accurate spectral classifications, the data remain affected by distance-dependent incompleteness. This effect is the result of the interplay between the increasing volume sampled with distance (scaling approximately as $r^2$ for a homogeneous three-dimensional distribution), and the progressive loss of faint objects due to the survey's detection limit (further exacerbated by interstellar extinction). As a result, the observed number of objects typically rises with distance up to a maximum and then declines, producing an artificial concentration at intermediate distances. Such selection effects can enhance the apparent prominence of structures surrounding the Sun, potentially giving rise to ring-like features whose inferred size depends sensitively on the survey's limiting magnitude and the homogeneity of the underlying classifications. In the case of OB stars, their high intrinsic luminosities push the peak in observed stellar counts to distances comparable to the scale-height of the Galactic disc, naturally leading to ring-like distributions instead of shells.

Another important effect arises from the use of the inverse of the parallax as a distance estimator \citep{Lurietal18}. This leads to the well-known Lutz–Kelker bias \citep{LutzKelk73, Smit03, Fran14}, which originates from the combination of parallax measurement uncertainties and the implicit assumption of a prior on the spatial distribution of sources. Because distance is a non-linear function of parallax, observational errors introduce asymmetric biases in the inferred distances, particularly at low signal-to-noise \citep{Weil25}. A simple way to illustrate the effect is to consider an idealized one-dimensional, homogeneous distribution of stars around the Sun, for which all distances are equally probable. In this case, however, the corresponding distribution of parallaxes is not uniform: large distances are compressed into a narrow range of small parallaxes, while nearby stars span a much wider range of larger parallax values. If parallax measurements are affected by homogeneous symmetric (e.g. Gaussian) uncertainties, this non-linear mapping leads to an asymmetric distribution of inferred distances when parallaxes are inverted. At small distances, where relative parallax uncertainties are low, the inferred distances closely trace the true distribution. At larger distances, however, small errors in parallax translate into large errors in distance, causing stars from a broad range of true distances to be scattered into lower measured distance intervals, where accurate measured distances are still present. At even larger distances the net flux of stars coming from lower and higher distances due to uncertainty slowly balances out. This results in a distortion of the recovered distance distribution, that showcases a peak in density at a certain distance. Moreover, if we shift to a more realistic three-dimensional distribution, where the number of stars increases with the square of the distance (for a spatially homogeneous distribution), there is a larger volume of distant stars that can be scattered into a given observed parallax compared to the number of nearby stars scattered out of it. As a result, the increase in dimensionality of space amplifies the bias that was already caused by the non-linearity of the transformation. Part of the problem is the surreptitious introduction of a uniform prior distribution of distances (which is in opposition to a uniform 3D distribution of stars, as it would mean that stars are preferentially clustered around the Sun, a highly non-Copernican prior assumption). For this reason Bayesian methods have usually been essential to reduce the effects of the Lutz-Kelker bias, which are difficult to correct when the relative uncertainties in the parallax grow large.

The distances derived for the ALS III catalogue of OB stars were the product of a Bayesian prior that included both a thin disc (appropriate for a kinematically cold young population) and a halo component with a small scale-height (to account for massive runaways). Combined with the high-precision astrometry of \textit{Gaia} DR3, this approach minimizes distance biases and does not produce spurious ring-like structures within the nearest kiloparsec. In contrast, earlier studies of the solar neighbourhood based on \textit{Hipparcos} data were more susceptible to the aforementioned biases, which may have artificially enhanced the appearance of Gould's Belt.

\subsection{Illusory kinematic patterns}

Due to the relativity of motion, a strong coupling can appear between the circular component of the peculiar motion of the Sun with respect to the LSR, $V_{\odot}$, and the rotational speed of the Galactic disk at the solar circle, $V_c$ (as both point in the same direction). This has divided the scientific community in two separate camps. On one side we have estimates of $V_{\odot} \approx 3 - 6$ km/s \citep{Mayo74, DehnBinn98, AumeBinn09, Goluetal13, Sysoetal18, Dingetal19}, while on the other we have studies that assert that $V_{\odot} \gtrsim 11$ km/s \citep{KerrLynd86, Schoetal10, BobyBajk18, ZbinSaha19}. The $V_{\odot}$ we choose has a considerable impact on the conclusions we gather from any study of the kinematics in the solar neighbourhood. When trying to define an accurate LSR, we aim to describe the average motion of the young stars in this region of the Galaxy, and from that to obtain a peculiar motion of the Sun and other nearby objects, so that no global drifts without a physical cause are induced by our choice of frame of reference. The nearest stars generally show the most prominent proper motions and these are also the most affected by the reflex motion of the Sun with respect to the surrounding matter, which makes them relevant in order to define the LSR. On the other hand basic geometry tells us that there are less stars nearby than farther away, meaning that local dynamics and low number statistics might bias our results.

So what is it? Is the Sun moving at higher speeds with respect to the nearby stars (which follow the overall trend of the disk), or is the Sun actually moving moderately slow and in fact is the nearby stars that are all moving as a single coherent structure in the opposite direction with respect to the bulk of the Galactic disk? To break this degeneracy we might use stars that are farther away from us, in order to examine whether the same solar reflex motion remains detectable in the peculiar velocities of stars once we have chosen a LSR. But the problem is that as we go farther away, the apparent contribution of the solar peculiar motion to the proper motions of stars diminishes rapidly and therefore higher relative uncertainties in the peculiar velocities are expected (because of higher relative uncertainties both in proper motion and in distances). In turn this means that observations of farther regions are not always helpful in discriminating between scenarios. Sometimes the best approach is to remind ourselves of the improbability of seeing all the surrounding medium having a peculiar motion almost opposite to that of the Sun, which is approximately what we find in the literature that deals with the bulk motion of Gould's Belt. By applying the Copernican principle to this situation we infer that we have probably been deceived by this bias in the past, before more accurate astrometry was available. In fact we can see examples of this happening across the literature. For example, when a global translational motion for Gould's Belt of $\sim 10$ km/s was detected in \citet{Boby04, Boby06}, the calculations were been made from the vantage point of the LSR defined by \citet{DehnBinn98}, which proposed a solar peculiar motion of $V_{\odot} = 5.20 \pm 0.62$ km/s. A decade afterwards, the same authors started using the much larger solar peculiar motion of $V_{\odot} = 12.24 \pm 0.47$ km/s, from \citet{Schoetal10}, when analysing the T Tauri stars of the Belt; the signature of a bulk motion for Gould's Belt then shrunk by a factor of two and even got to the point of practically vanishing \citep{Boby16, BobyBajk20}.

\begin{figure}
    \centering
    \includegraphics[width=0.4\textwidth, trim=15 39 36 37, clip]{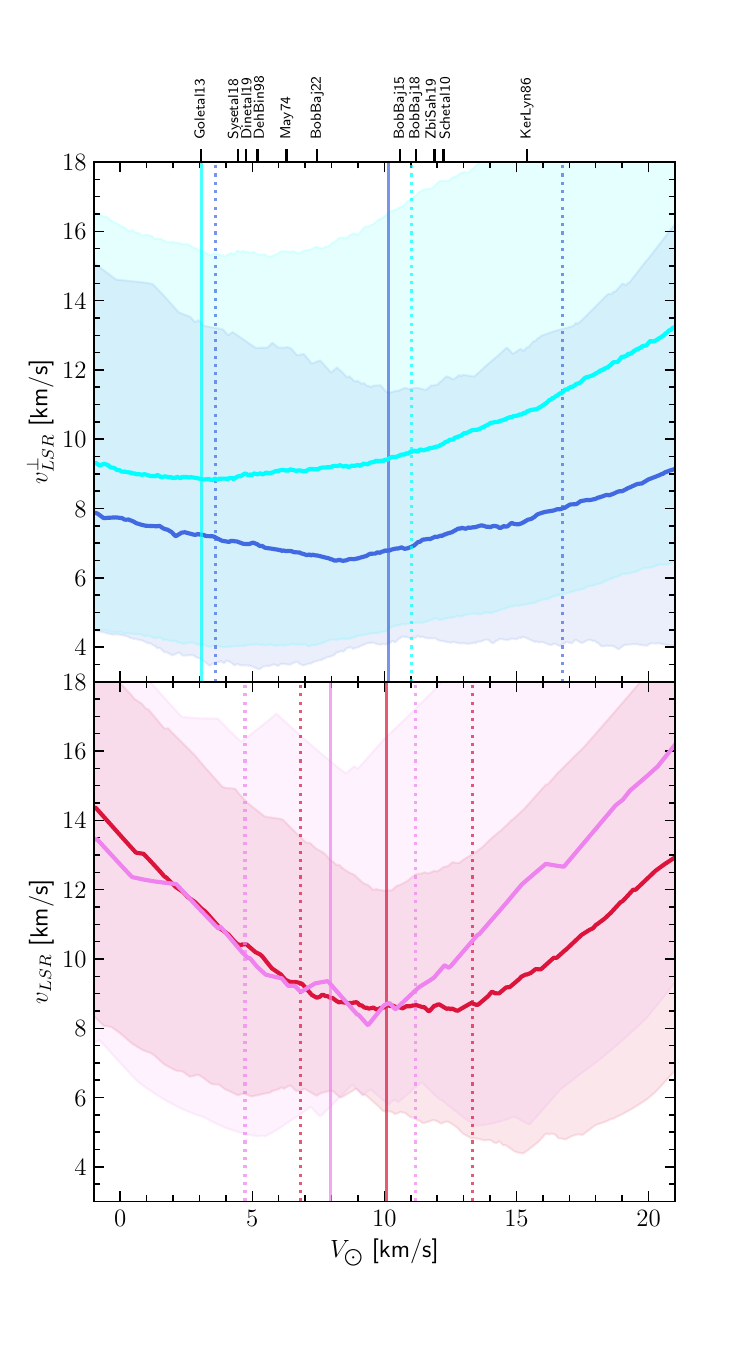}
    \caption{Median peculiar velocities calculated as a function of $V_{\odot}$ when defining a LSR. The filled areas represent the dispersion bounded by the $16$th and $84$th percentiles. (\textit{Top panel}) Massive stars in the Belt (dark blue) and outside the Belt (cyan). (\textit{Bottom panel}) Young clusters in the Belt (crimson) and outside the Belt (pink). The vertical continuous and dotted lines represent the resulting $V_{\odot}$ values of the $\chi^2$ minimization procedure and their uncertainties respectively. Values of $V_{\odot}$ taken from the literature are marked at the top of the the figure with abreviations following [BobBaj15] \citet{BobyBajk15}, [BobBaj18] \citet{BobyBajk18}, [BobBaj22] \citet{BobyBajk22}, [DehBin98] \citet{DehnBinn98}, [Dinetal19] \citet{Dingetal19}, [Goletal13] \citet{Goluetal13}, [KerLyn86] \citet{KerrLynd86}, [May74] \citet{Mayo74}, [Schetal10] \citet{Schoetal10}, [Sysetal18] \citet{Sysoetal18} and [ZbiSah19] \citet{ZbinSaha19}.}
    \label{fig:fig4}
\end{figure}

We can illustrate this point practically with our own data. In Fig. \ref{fig:fig3} we calculated the peculiar velocities for stars and clusters in the Gould's Belt as a function of Galactic longitude by considering the LSR defined in \citet{Schoetal10} and the rotation curve in \citet{Sofu20}. For the OB stars (top panel in Fig. \ref{fig:fig3}) we lack accurate radial velocities so the peculiar motion presented is the modulus of the component tangential to the celestial sphere, $v_{LSR}^{\perp}$. For the young clusters (bottom panel in Fig. \ref{fig:fig3}) the modulus of the total peculiar velocity, $v_{LSR}$, can be computed accurately as the combined radial velocities of all the stars within each cluster constrain this value. For massive stars in the Belt we find typical tangential peculiar velocities of $v_{LSR}^{\perp} = 7.2_{-3.1}^{+4.5}$ km/s, while for the clusters we find total peculiar tangential velocities of $v_{LSR} = 8.6_{-3.3}^{+4.0}$ km/s. To compute these values we have taken the median to avoid the effect of some kinematic outliers present in the OB sample, as there are walkaway and runaway stars still present on it. Correspondingly we measured the dispersion in these velocities with the range defined by the $16$th and $84$th percentiles. No cluster in our sample appears to have a significantly larger peculiar motion. The value for the clusters is slightly larger to that of the massive stars due to the fact that the peculiar velocity also includes the radial component. The interesting results are shown in the \href{https://youtu.be/LSYs4zPPMDQ}{online animated version} of Fig. \ref{fig:fig3}, where the peculiar velocities are calculated as we continuously change the $V_{\odot}$ component of the peculiar motion of the Sun with respect to the LSR (we are interested in this component since this is the one that can easily couple with the circular velocity of the LSR, $V_c$). Since the radial component is lacking in the peculiar velocities shown for the OB stars, a bulk translational relative motion of a structure covering the whole sky would appear as an oscillation in the peculiar tangential velocities across Galactic longitude. This is behaviour is not apparent in the LSR defined by \citet{Schoetal10}, but clearly emerges as we shift to different $V_{\odot}$ values. Another observation that can be made is that the peculiar velocities of both samples are simultaneously minimized for values of $V_{\odot}$ that are close to the ones defined by \citet{Schoetal10}, implying that these values are correctly anchoring the LSR to our immediate surroundings. We can quantify this by performing a $\chi^2$ minimization of the peculiar velocities as a function of $V_{\odot}$ (see Fig. \ref{fig:fig4}). The optimal LSR (defined as the one minimizing peculiar motions) is found at $V_{\odot} = 10.2 \pm 6.6$ km/s for the Gould Belt massive stars and $V_{\odot} = 10.1 \pm 3.2$ km/s for the young clusters, in good agreement with each other and with the value $V_{\odot} = 12.24 \pm 0.47$ km/s reported by \citet{Schoetal10}. This agreement is expected given the relatively large uncertainties in our measurements. While refining the LSR is beyond the scope of this paper, the key point is that even simple methods applied to reliable astrometry reproduce values that are found in the dedicated literature. But this could still mean that the LSR has been anchored to the Gould Belt and that a different LSR might be defined when looking to the population outside of it. For this reason we also analyse a sample of outside-the-Belt massive stars and clusters, defined by the distance range between $600$ pc (clearly outside of Gould's Belt) to $1200$ pc (where the astrometry is still generally highly reliable). For the nearby OB stars, we find $V_{\odot} = 3.1 \pm 7.9$ km/s, a value that even if it is compatible with that obtained for the Gould's Belt sample is of limited significance due to the large uncertainty. The reason for this being the dispersion in peculiar velocities driven by the presence of runaway stars in our sample (we will identify these in a future paper) and the presence, inside the $1.2$ kpc radius, of kinematically coherent structures like the Cepheus Spur \citealt{Pantetal21}, that could slightly bias the results. For young clusters outside the Belt, since radial velocities are available, we obtain a more robust $V_{\odot} = 8.0 \pm 3.2$ km/s. The resulting difference in $V_{\odot}$ between Belt and non-Belt clusters is $2.1 \pm 4.6$ km/s, indicating no detectable kinematic signature distinguishing the Gould Belt from the surrounding population. This dataset (more accurate and complete than what was available during the \textit{Hipparcos} era) can rule out a bulk translational peculiar motion as large as the one reported by \citet{Boby04a, Boby06}. But if the kinematics of the Belt are no different than those found in the rest of the solar neighbourhood how was such motion detected in pre-\textit{Gaia} works? The answer is a combination of different completeness as we move to samples farther from the Sun, and as stated before a larger relative uncertainty in the peculiar velocities as father distances.

\section{Conclusions}
The analysis presented here suggests that Gould's Belt does not exist as a unique coherent dynamical structure. Its apparent expansion and rotation arise from the superposition of a few nearby young cluster families, whose slight relative motions, when viewed from the Sun, mimic the kinematic signatures (expansion and rotation) historically attributed to the Belt. The reported bulk motion of the Belt is largely an artifact of adopting outdated or inaccurate values for the Sun's peculiar motion relative to the LSR. Although the spatial distribution of young massive stars clearly exhibits a ring-like pattern, we show that this configuration is transient and results from the chance alignment of independent dynamical populations. This ring appearance is thus expected to quickly dissolve within the next $\sim 30$ Myr. Finally, the commonly cited $\sim 15^\circ - 20^\circ$ inclination of Gould's Belt is based on the fact that the anti-central arc of the Belt is actually part of the Radcliffe Wave (which goes from higher latitudes in Cepheus and Perseus to its lowest point in Orion) combined with the high latitude of Sco-Cen, rather than evidence for a physically distinct inclined plane connecting all of these populations. Taken together, these findings strongly indicate that Gould's Belt exists only as a modern tridimensional asterism rather than a physically meaningful Galactic feature, and that a new, more nuanced description of the structure and evolution of the solar neighbourhood supersedes it.

\begin{acknowledgements}
    Co-Funded by the European Union (ERC, ISM-FLOW, 101055318). Views and opinions expressed are, however, those of the author(s) only and do not necessarily reflect those of the European Union or the European Research Council Executive Agency. Neither the European Union nor the granting authority can be held responsible for them. Authors acknowledge funding from the European Research Council (ERC) under the European Union’s Horizon 2020 research and innovation programme (Grant agreement No. 851435). This work has made use of data from the European Space Agency (ESA) mission Gaia (\href{https://www.cosmos.esa.int/gaia}{https://www.cosmos.esa.int/gaia}), processed by the Gaia Data Processing and Analysis Consortium (DPAC, \href{https://www.cosmos.esa.int/web/gaia/dpac/consortium}{https://www.cosmos.esa.int/web/gaia/dpac/consortium}). Funding for the DPAC has been provided by national institutions, in particular the institutions participating in the Gaia Multilateral Agreement.
\end{acknowledgements}

\bibliographystyle{aa}
\bibliography{general}

\end{document}